\documentclass[aps,prl,showpacs]{revtex4}

\usepackage{graphicx}
\usepackage{hyperref}
\usepackage{amsfonts}
\usepackage{amsbsy}
\usepackage{amssymb}
\usepackage{xcolor}
\usepackage[normalem]{ulem}
 \usepackage{float}
 \usepackage{multirow}
 
\begin{document}


\title{Approaching a large deviation theory for complex systems} 

\author{Ugur Tirnakli$^{1}$}
\email{ugur.tirnakli@ege.edu.tr}
\author{Constantino Tsallis$^{2,3,4}$}
\email{tsallis@cbpf.br}
\author{Nihat Ay$^{3,5}$}
\email{Nihat.Ay@mis.mpg.de}

\affiliation{ $^1$Department of Physics, Faculty of Science, 
Ege University, 35100 Izmir, Turkey \\
$^2$Centro Brasileiro de Pesquisas Fisicas 
and National Institute of Science and Technology for Complex Systems \\
\mbox{Rua Xavier Sigaud 150, Rio de Janeiro 22290-180, Brazil} \\
 $^3$ Santa Fe Institute, 1399 Hyde Park Road, Santa Fe, 
 New Mexico 87501, USA \\
 $^4$ Complexity Science Hub Vienna, Josefst\"adter Strasse 
 39, 1080 Vienna, Austria \\
 $^5$Max Planck Institute for Mathematics in the Sciences
Inselstrasse 22, 04103 Leipzig, Germany
 }

\date{\today}

\begin{abstract}
The standard Large Deviation Theory (LDT) is mathematically illustrated by the Boltzmann-Gibbs 
factor which describes the thermal equilibrium of short-range-interacting many-body Hamiltonian systems, 
the velocity distribution of which is Maxwellian.
It is generically applicable to systems satisfying the Central Limit Theorem (CLT). 
When we focus instead on stationary states of typical complex systems 
(e.g., classical long-range-interacting many-body Hamiltonian systems, such as self-gravitating ones), 
the CLT, and possibly also the LDT, need to be generalised. 
Specifically, when the $N\to\infty$ attractor ($N$ being the number of degrees of freedom) 
in the space of distributions is a $Q$-Gaussian (a nonadditive $q$-entropy-based generalisation 
of the standard Gaussian case, which is recovered for $Q=1$) related to a $Q$-generalised CLT, 
we expect the LDT probability distribution to asymptotically approach 
a power law. Consistently with available strong numerical indications for probabilistic models, 
this behaviour possibly is that associated to a $q$-exponential 
(defined as $e_q^x\equiv\left[1+(1-q)x\right]^{1/(1-q)}$, which is the generalisation of the 
standard exponential form, straightforwardly recovered for $q=1$); $q$ and $Q$ are expected to be simply connected, 
including the particular case $q=Q=1$. 
The argument of such $q$-exponential would be expected to be proportional to $N$, analogously to 
the thermodynamical entropy of many-body Hamiltonian systems. 
We provide here numerical evidence supporting the asymptotic power-law 
by analysing the standard map, the coherent noise model for biological extinctions and earthquakes, 
the Ehrenfest dog-flea model, and the random-walk avalanches. 
For the particular case of the strongly chaotic standard map, we numerically verify (below $5\%$ error bar) 
the validity of the asymptotic exponential behavior predicted by the usual LDT once the initial transient elapses 
typically beyond $N \simeq 3 \times 10^6$. Analogously, for the standard map with vanishing Lyapunov exponent, 
we provide numerical evidence (below the same error bar) for the asymptotic validity of the $q$-exponential 
behavior once the initial transient elapses typically beyond $N \simeq 2 \times 10^5$.
\end{abstract}


\maketitle

\section{Introduction}

Boltzmann-Gibbs (BG) statistical mechanics yields various important relations. 
Still, it is fair to consider as its most important fingerprints the Maxwellian distribution 
of velocities and the exponential distribution of energies (BG weight or BG factor) \cite{statphys1,statphys2}. 
These behaviours correspond mathematically to the Central Limit Theorem (CLT) \cite{CLT1,CLT2} which leads, 
when the number $N$ of involved random variables increases indefinitely, to convergence towards 
Gaussian distributions, and to the Large Deviation Theory (LDT) \cite{LDT1,LDT2,Touchette2009} which 
characterises the speed at which Gaussians are approached while $N$ increases. 
To be more precise, the BG distribution $p_{BG}$ associated with a many-body Hamiltonian ${\cal H}_N$ 
at thermal equilibrium is given by $p_{BG} \propto e^{-\beta {\cal H}_N}$ whenever ${\cal H}_N$ 
includes short-range interactions or no interactions at all. 
We may then write that $p_{BG} \propto e^{-[\beta {\cal H}_N/N]N}$, where, consistently with thermodynamics, 
$[\beta {\cal H}_N/N]$ is an intensive quantity. The corresponding LDT statement concerns 
the probability $P_N(Y_N/N >z) \in [0,1]$ of the random variable $Y_N/N$ taking values larger than 
a fixed value $z \in\Re$ for increasingly large values of $N$. Under the hypothesis of probabilistic 
independence, or similar settings, we expect $P_N(Y_N/N >z) \approx e^{- r_1(z)N}$, where the 
{\it rate function} $r_1$ equals a {\it relative entropy per particle}. 
Therefore $r_1(z) N$ plays the role of the thermodynamic total entropy which, consistently with the Legendre 
structure of classical thermodynamics, is extensive, i.e., $r_1(z)N\propto N$ ($N\gg 1)$.

Here we focus on systems with nonlocal space-time correlations by generalising 
the BG theory \cite{Tsallis1988,Tsallisbook}. 
The basis of this generalisation consists in optimising entropies which differ 
from $S_{BG}=-k\sum_i p_i \ln p_i$, such as $S_q=k \frac{1-\sum_ip_i^q}{q-1}$ with 
$q \in {\cal R}$, and $S_1=S_{BG}$ (see \cite{JizbaKorbel2019} for an interesting discussion of the admissibility 
of such generalisation on statistical inference grounds). If $A$ and $B$ are two probabilistically independent 
systems, we straightforwardly verify that $S_q(A+B)/k= S_q(A)/k+S_q(B)/k+(1-q)[S_q(A)/k][S_q(B)/k]$. 
In other words, $S_q$ is nonadditive for $q \ne 1$ whereas $S_{BG}$ is additive. The optimisation of 
$S_q$ with simple constraints yields a probability distribution 
$p_q\propto e_q^{-\beta_q {\cal H}_N}$, where $\beta_q$ plays the role of an inverse temperature and 
$e_q^z \equiv [1+(1-q)z]_+^{1/(1-q)}$ with $e_1^z=e^z$ and 
$[\dots]_+=[\dots] $ if $[\dots] >0$, and zero otherwise (whose inverse function is the $q$-logarithm 
defined as $\ln_q z \equiv (z^{1-q}-1)/(1-q)$). The form $e_q^{-a\,z^2}$ ($a>0$) is usually 
referred to as $q$-Gaussian. This $q$-generalised statistical mechanics ({\it nonextensive statistical mechanics} 
or {\it $q$-statistics} for short)  typically tackle with long-range-interacting Hamiltonian systems, e.g., 
self-gravitating systems \cite{TaruyaSakagami2003}, violently relaxing systems \cite{CampaChavanisGiansantiMorelli2008}, ionic crystals \cite{ioniccrystals1,ioniccrystals2}, 
among other nontrivial systems such as cold atoms in dissipative lattices \cite{Renzoni,LutzRenzoni}, 
granular matter \cite{Combe}, high-energy collisions of elementary particles \cite{WongWilk}, 
overdamped systems like type-II superconductors \cite{Andrade}, matter-antimatter astrophysical 
observations \cite{Beck}, stellar physics \cite{Freitas2019}, theory of finances \cite{Borland}, complex networks \cite{OliveiraBritoSilvaTsallis2021}.
The associated distributions of velocities appear to be $Q$-Gaussians with $Q > 1$ 
(see, for instance, \cite{AnteneodoTsallis1998,CirtoRodriguezNobreTsallis2018} 
for the $\alpha$-XY ferromagnet, \cite{RodriguezNobreTsallis2019} for the $\alpha$-Heisenberg ferromagnet, 
and \cite{ChristodoulidiTsallisBountis2014,BagchiTsallis2016} for the $\alpha$-Fermi-Pasta-Ulam model), 
with $Q$ approaching unity when the range of the interactions approaches the short-range regime. 
This corresponds mathematically to a $Q$-generalised Central Limit Theorem ($Q$-CLT) which leads, 
when the number $N$ of strongly correlated random variables increases to infinity, to a convergence on a 
$Q$-Gaussian distribution. Sufficient conditions for the $Q$-CLT to hold are already 
available \cite{UmarovTsallisSteinberg2008} (see also \cite{UmarovTsallisGellMannSteinberg,umarov}) but the 
necessary conditions for a $Q$-CLT remain as a challenge. 
A BG approach of long-range-interacting Hamiltonian systems can be seen 
in \cite{CampaDauxoisFanelliRuffo2014}. However, the fact that various thermostatistical quantities are 
computable within the BG theory by no means guarantees that the theory correctly handles the many-body dynamics 
of those systems at any experimentally accessible time for any experimentally accessible size of the system. 
Analytical computability is necessary but not sufficient. Indeed, in the various anomalous systems listed above, 
the BG approach poorly fits 
reality \cite{ioniccrystals1,ioniccrystals2,Renzoni,LutzRenzoni,Combe,WongWilk,Andrade,Beck,Freitas2019,Borland,OliveiraBritoSilvaTsallis2021,AnteneodoTsallis1998,CirtoRodriguezNobreTsallis2018,RodriguezNobreTsallis2019,ChristodoulidiTsallisBountis2014,BagchiTsallis2016}. 
 
Within $q$-statistics we have, for the total energy of the system at its stationary, 
or quasi-stationary, state, $p_q \propto e_q^{-\beta_q {\cal H}_N}$, with ${\cal H}_N$ being 
super-extensive, i.e., not proportional to $N$, consistently with long-range interactions. 
For say two-body (attractive) interactions decaying like $1/(distance)^\alpha$ ($\alpha \in [0,\infty)]$) 
within a $d$-dimensional system, we may rewrite 
$p_q \propto e_q^{-[(\beta_q \tilde{N}) ({\cal H}_N/N\tilde{N})]N}$ where 
$\tilde{N} \equiv \frac{N^{1-\alpha/d}-1}{1-\alpha/d}$ is, for $N$ increasingly large, 
constant for $\alpha/d>1$ (short-range), increases like $N^{1-\alpha/d}$ for $0 \le \alpha/d <1$ (long-range), 
and increases like $\ln N$ for $\alpha/d=1$.  
Notice that both  $ (\beta_q \tilde{N})$ and $({\cal H}_N/N\tilde{N})$ are intensive quantities 
(see details in \cite{TsallisCirto2013} and references therein). The desirable mathematical correspondence 
would of course be to have a $q$-Large Deviation Theory ($q$-LDT) with a probability 
$P(N, Y/N >z)\approx e_q^{- r_q(z)N}$, where the {\it rate function} $r_q$ would once again equal some 
sort of {\it relative nonadditive entropy $S_q$ per particle}. 
Therefore $r_q(z) N$ is expected to play the role of the total system thermodynamic entropy which, 
as before, should be extensive, i.e., $\propto N$ ($N\gg 1)$. Naturally, in order to unify all the above 
situations, we expect $q=f(Q)$, $f(Q)$ being a smooth function which satisfies $f(1)=1$, 
thus recovering the usual LDT.
 
The above $q$-LDT scenario has already been numerically verified for a purely probabilistic model with strong 
correlations \cite{RuizTsallis2012,Touchette2012,RuizTsallis2013}. 
In the present paper we follow along those lines and focus on the possible emergence of the same type of 
probability for four well known dynamical models, namely the standard map, the coherent noise model for biological 
extinctions and earthquakes, the Ehrenfest dog-flea model, and the random walk avalanches. 
The Ehrenfest model is a genuine $N$-body problem, whereas  what plays the role of $N$ in the 
other three models is the number of successive iterations.
In these four models we numerically verify that, in the space of probability distributions, 
convergence towards $Q$-Gaussians indeed occurs when summing
increasingly large number $N$ of random variables. 
Here, we focus on the speed at which these $Q$-Gaussians are attained when $N$ increases. 
In other words, we provide dynamical examples approaching the mathematical development of a $q$-LDT. 

Concerning the standard LDT, several nontrivial calculations, either analytical or numerical, 
are available in the literature, where various classes of systems are focused 
on \cite{ref1,ref2,ref3,ref4,ref5,ref6,ref7,ref8}. Such systems may be either stochastic or Hamiltonian ones. 
For example, rogue waves, Fermi-Pasta-Ulam-Tsingou chains, population dynamics including birth-death processes, 
deterministic systems such as the Lorentz gas, Markov dynamics involving both symmetric and asymmetric exclusion 
processes, glass models exhibiting various dynamical phenomena such as super-Arrhenius temperature dependence 
of characteristic times, non-exponential relaxation, spatially heterogeneous dynamics, transport decoupling, 
ageing  and memory effects.\\

\section{Results}
We address now a low-dimensional conservative 
system, namely the {\it standard map} \cite{chirikov1}:
\begin{eqnarray}
 \label{eq:stdmap}
 U_{i+1} &=& U_i - K \sin X_i  \nonumber \\
 X_{i+1}&=& X_i+ U_{i+1} \;\;\;\;(K \ge 0) \,,
\end{eqnarray}
($U$ and $X$ are taken as modulo $2\pi$). 
This is a highly paradigmatic system in the study of Hamiltonian 
low-dimensional maps \cite{zaslavsky91,zaslavsky05} and 
has been analyzed deeply in the context of physical 
applications \cite{Izraelev1980,Petrowsky1986,Benvenuto1994} 
as well as mathematical aspects in the theory of dynamical 
systems \cite{Greene86,Aubry}. Its applications include particle confinement in magnetic traps, 
particle dynamics in accelerators, 
comet dynamics, ionisation of Rydberg atoms and electron magneto-transport.

This map is integrable for $K=0$ and non-integrable otherwise. If $K\ll 1$ the phase space 
of the system is dominated by the stability islands. As $K$ increases, chaotic behaviour starts 
to set in for a tiny portion of the phase space. There is a critical $K$ value ($K_c\approx0.97$) 
below which the chaotic regions in the phase space do no communicate, and there is a unique chaotic 
sea for $K>K_c$. 
If $K\gg 1$ the chaotic sea dominates a large portion of the phase space and for some $K$ 
values the stability islands become virtually invisible \cite{Srivastava,Tomsovic}. 
On the other hand, it also appears that, as $K\to\infty$, there will be islands for a residual set of $K$ 
values \cite{Duarte}.
For the phase space portions with stability islands, the system exhibits weak chaotic dynamics, 
characterised by zero local Lyapunov exponent. Instead, for the portions of chaotic sea, 
it exhibits strong chaotic dynamics with a positive Lyapunov value. 
Now let us concentrate on the $X$ variable of the map and define a random variable as the sum 
of iterates of the map: 
\begin{equation}
 \label{eq:y}
 Y_N:=\sum_{i=1}^N (X_i - \langle X_i \rangle )\, ,
\end{equation}
where $\langle \cdots \rangle$ represents the expectation value and we approximate this by 
sampling over a large number of $M$ initial conditions taken randomly from uniform distribution in 
$[0,2\pi]$, that is,
\begin{equation}
\label{eq:average}
 \langle X_i \rangle \approx \frac{1}{M}\sum_{j=1}^M  X_i^{(j)} \, .
\end{equation}
For large $M$, this will converge to a number $\mu$ independent of $i$. Therefore, 
Eq.~(\ref{eq:y}) translates to 
\begin{equation}
\label{eq:yN}
  Y_N= \sum_{i=1}^N X_i -N\mu \, ,
\end{equation}
which enables us to create a sequence of data points centered around zero from a deterministic 
dynamical system.


\noindent
Let us focus on the $K=10$ case. The CLT basically states that the sum of $N$ independent identically 
distributed (i.i.d.) random variables (appropriately centred and rescaled), converges onto a 
Gaussian distribution in the limit $N\to\infty$ (property that is also currently referred to by 
saying that a Gaussian is the {\it attractor in the space of distributions}, not to be confused 
with the {\it dynamical attractor in phase space}, a completely different concept). 
Although it is evident that the iterates of a deterministic dynamical system would never be 
completely independent, one can still prove CLTs if the i.i.d. assumption is replaced by the property that 
the system is strongly mixing \cite{mackey06,bill}, which is guaranteed by strong chaos. 
Therefore, in the context of the standard map, one would expect the usual CLT to hold for $K=10$. 
More precisely, $\sqrt{N}(Y_N/N)$ should (weakly) converge, for $N\to\infty$, to a variable with a 
Gaussian distribution  \cite{ugur16}: 

\begin{equation}
\lim_{N\rightarrow\infty} P_N\left(a\leq \sqrt{N}Y_N/N\leq b\right) 
=\int_a^b p(y;\sigma) dy \, 
\end{equation}
or equivalently

\begin{equation}
\label{PN}
\lim_{N\rightarrow\infty} P_N\left(\sqrt{N}Y_N/N\geq z\right) 
=\int_z^{\infty} p(y;\sigma) dy \, ,
\end{equation}
where the probability density is given by
\begin{equation}
 \label{density}
p(y; \sigma)
=\frac{1}{\sigma \sqrt{2 \pi }}
\exp\left[-\frac{1}{2}\left(\frac{y}{\sigma}\right)^2 \right] \, .
\end{equation}
We remind that the prefactor $\sqrt{N}$ before $Y_N$ emerges in order to have gradual data 
collapse for increasingly large $N$.

\begin{figure}
\centering
\includegraphics[scale=0.187,angle=0]{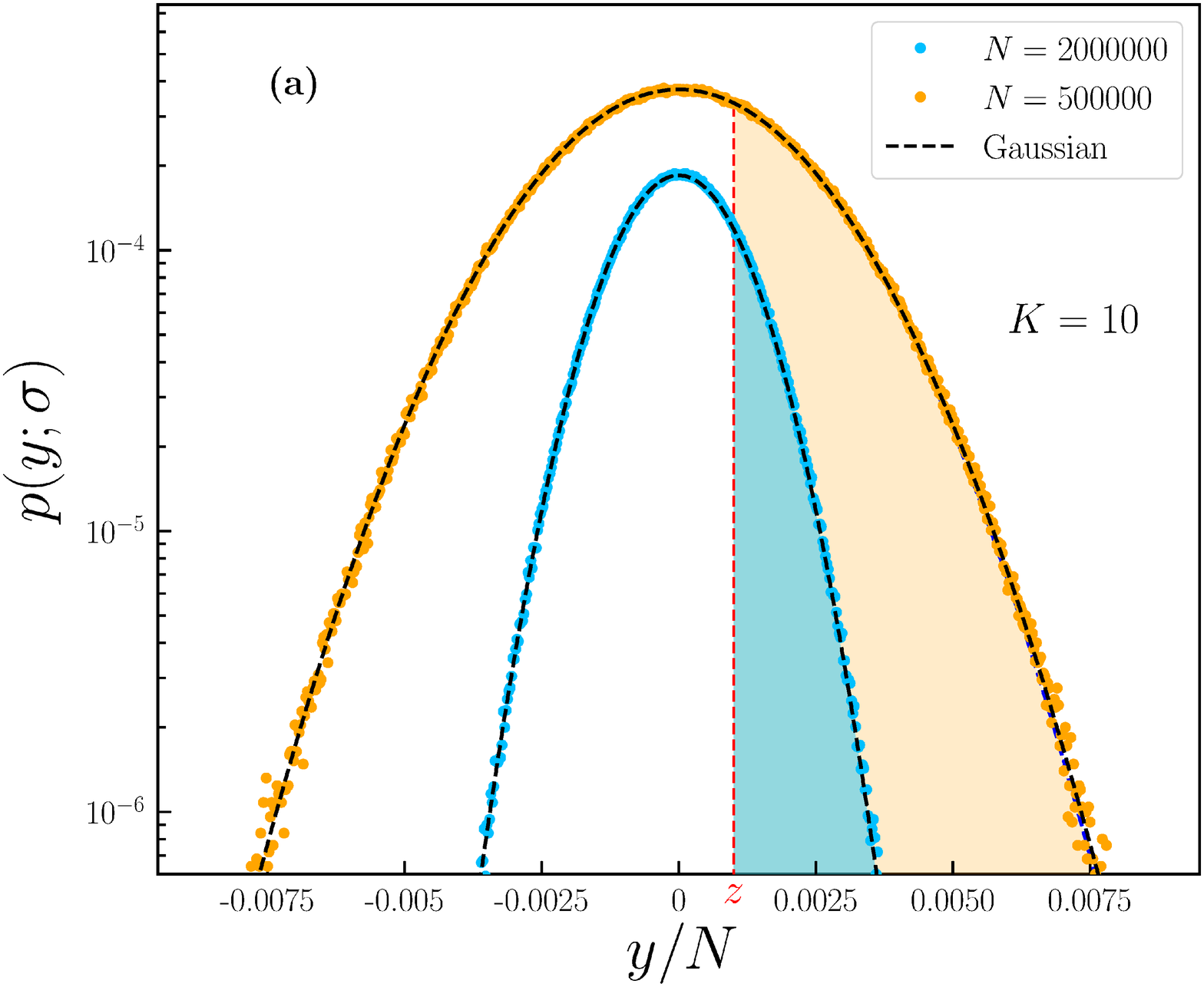}
\includegraphics[scale=0.187,angle=0]{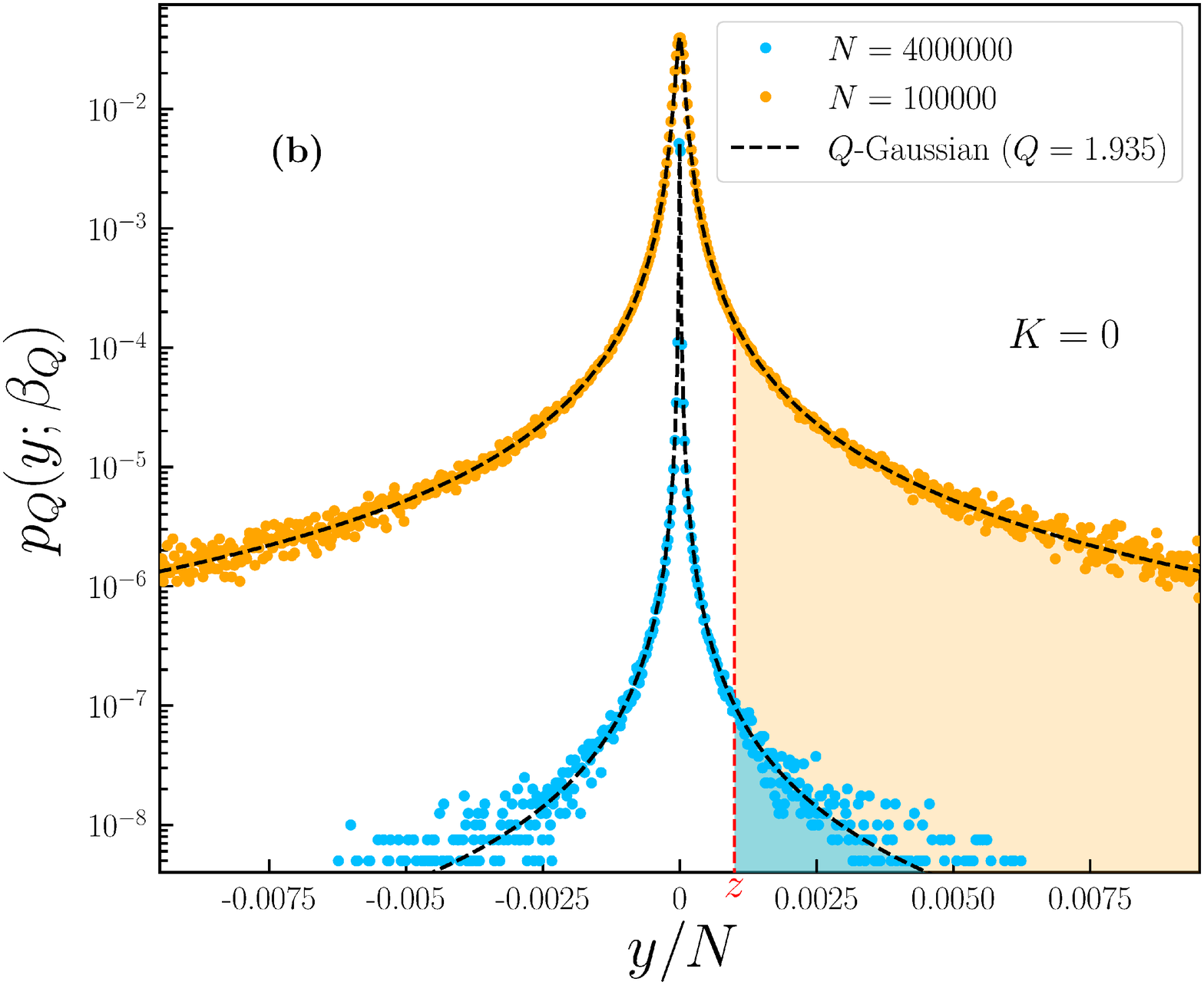}
\includegraphics[scale=0.187,angle=0]{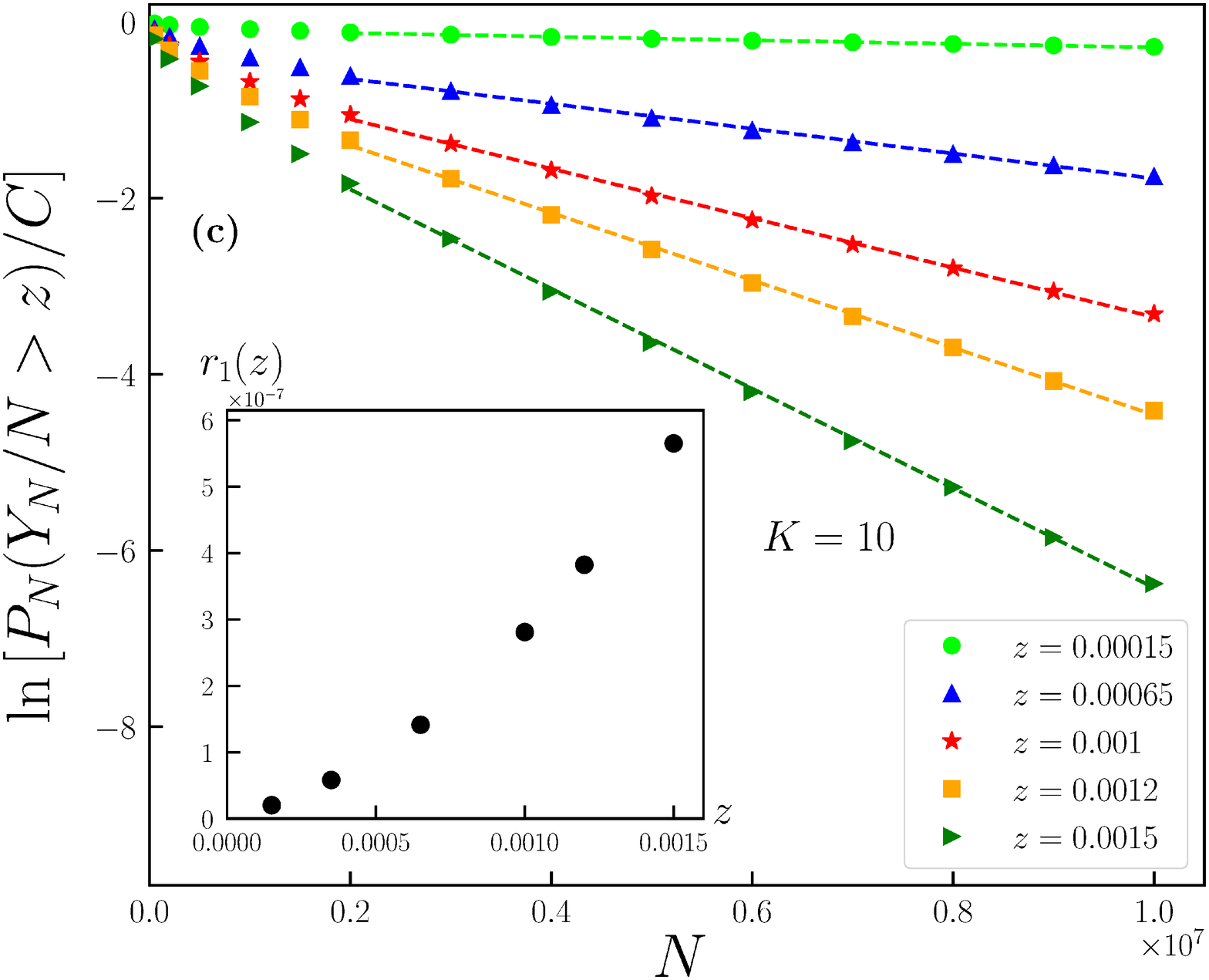}
\includegraphics[scale=0.187,angle=0]{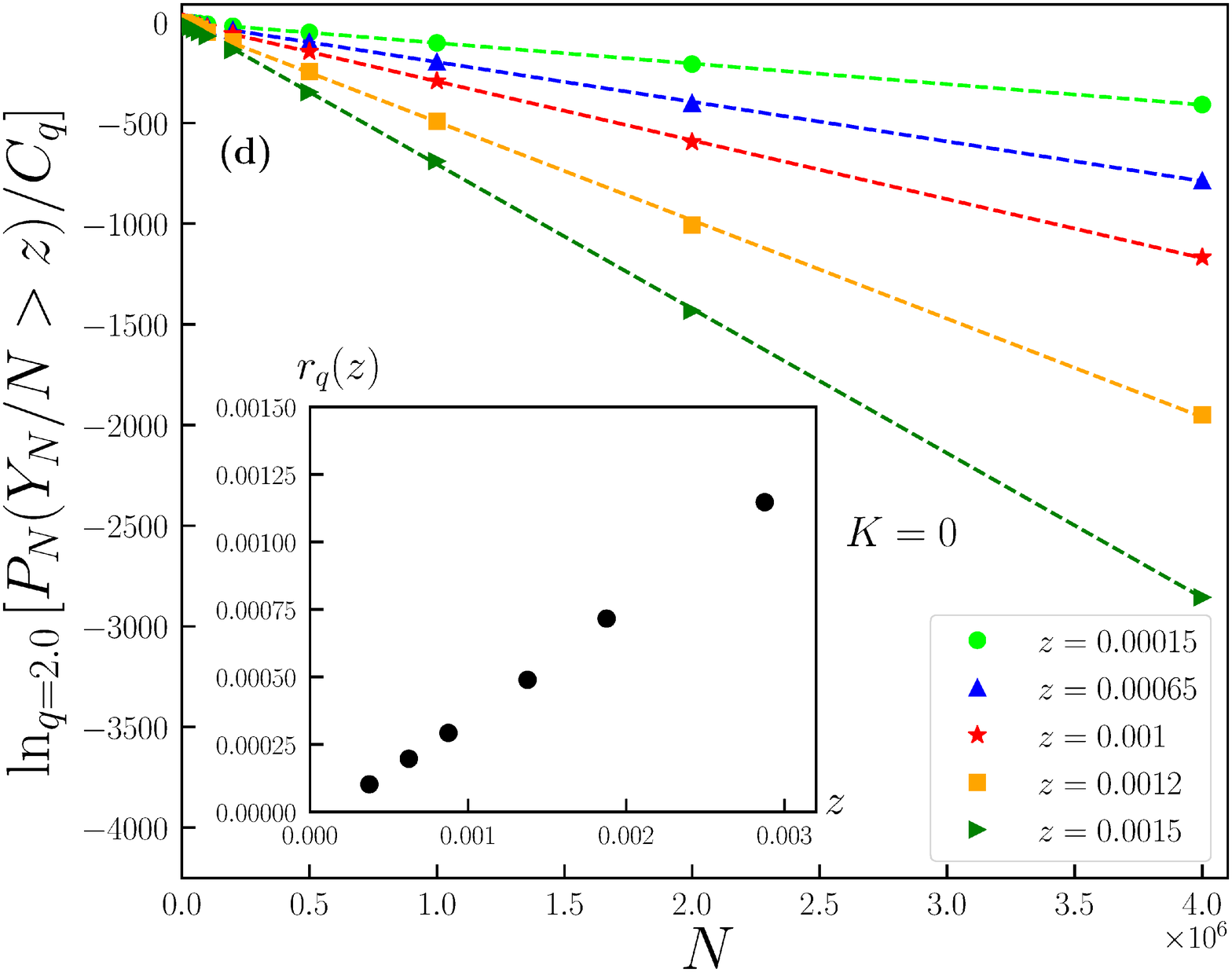}
\includegraphics[scale=0.185,angle=0]{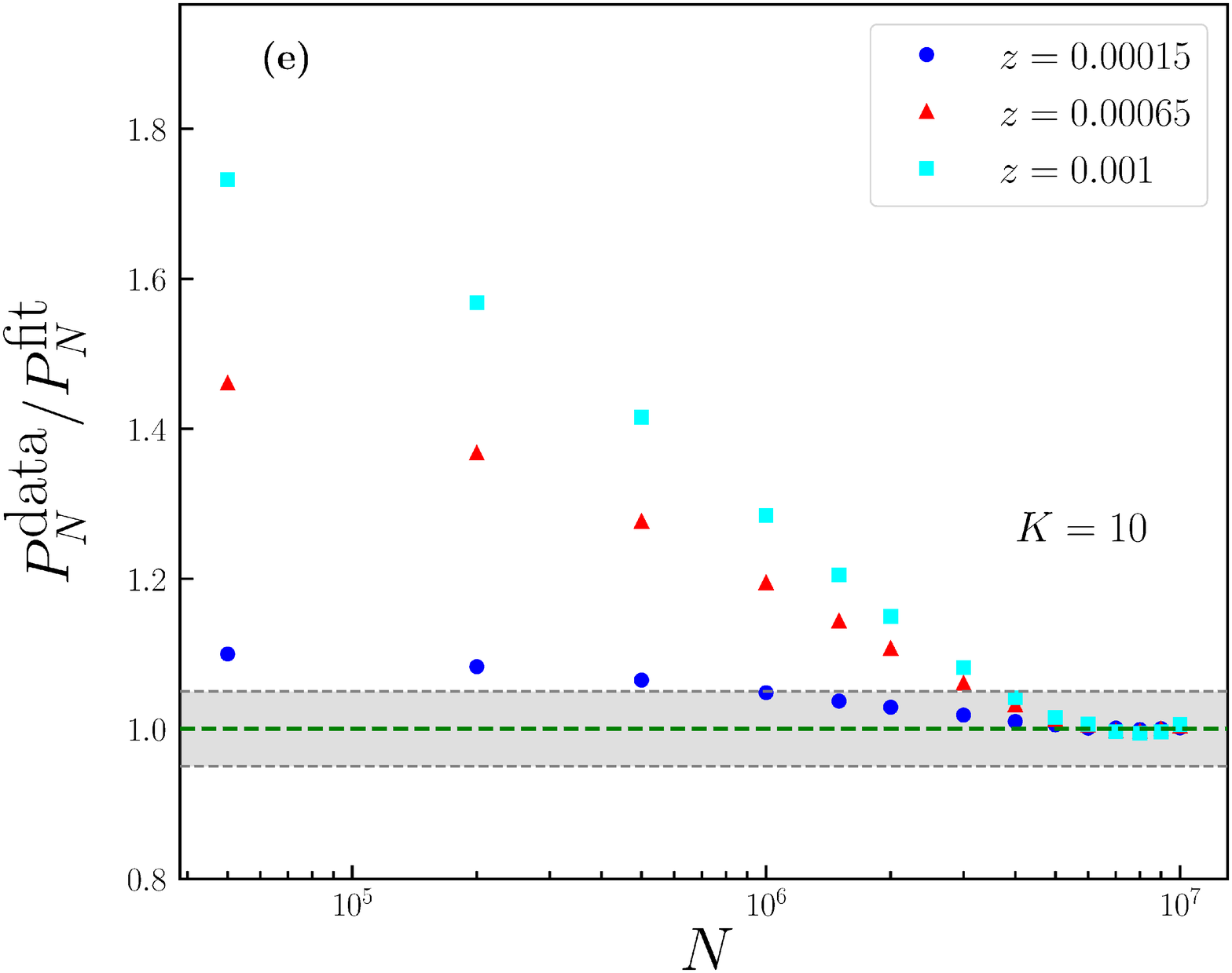}
\includegraphics[scale=0.185,angle=0]{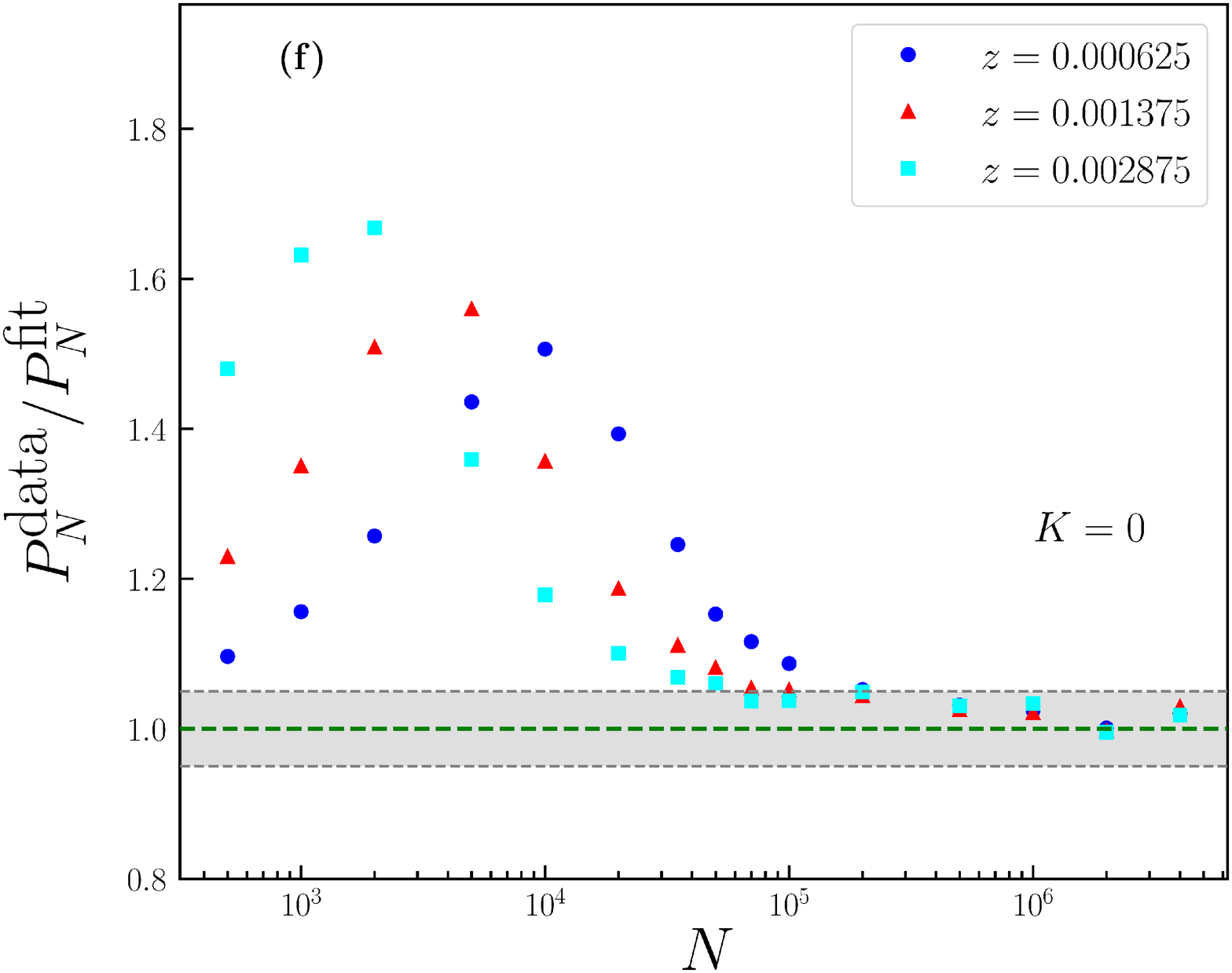}
\caption{\label{fig:Fig1} {\small (Color online) 
The probability density function of the standard map for two representative values of $N$ for 
(a)~$K=10$ and (b)~$K=0$. 
If we multiply both the ordinate and the abscissa by $\sqrt{N}$ in (a) and by $N^{\gamma}$ 
with $\gamma \simeq 0.65$ in (b), the present data collapse onto a single Gaussian 
($Q$-Gaussian with $Q\simeq 1.935$) for $K=10$ ($K=0$). 
A typical value $z>0$ is indicated as well. 
In (c) and (d), we see respectively that the large deviation probability $P_N(Y_N/N > z)$ 
asymptotically decays with $N$ exponentially for $K=10$ and 
as a power-law (possibly $q$-exponentially with $q=2.0$) for $K=0$. 
The slopes provide the rate function $r_q(z)$ as shown in the Insets. 
Finally in (e) and (f), we represent the ratio between the data and fitted function 
for $K=10$ and $K=0$ respectively. The discrepancy bars at $1 \pm 0.05$ are indicated in dotted lines.} }
\end{figure}

For $K=10$, the dynamics displays strong chaos within the full (or nearly full) phase space.   
This implies the convergence of $Y_N/N$ to the Gaussian distribution. 
In Fig.~\ref{fig:Fig1}a, the distribution of $Y_N/N$ is shown for two representative values of $N$. 
The Gaussian shape is evident and a representative $z>0$ value is also indicated in the figure. 

\begin{figure}
\centering
\includegraphics[scale=0.24,angle=0]{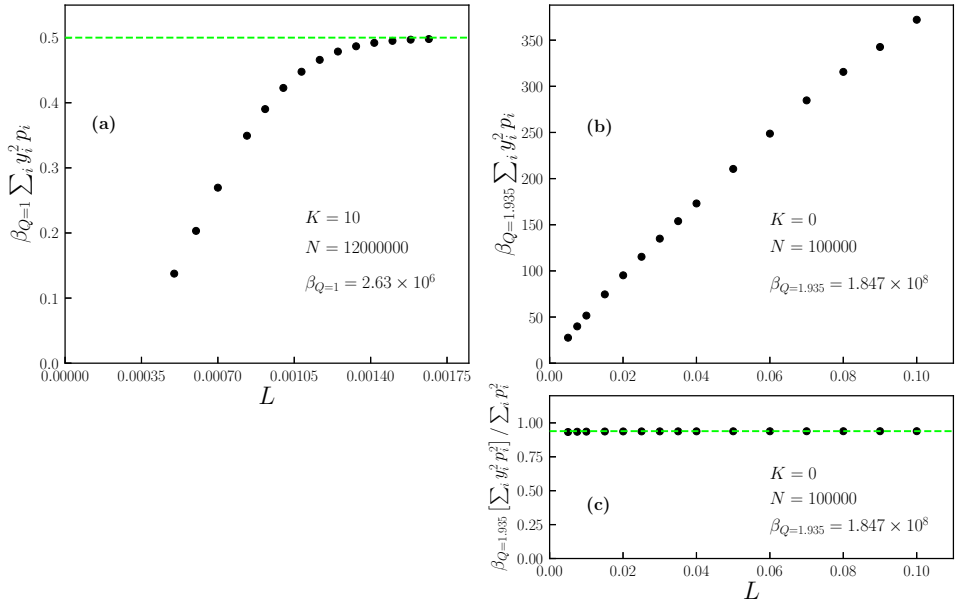}
\caption{\label{fig:Figplus} {\small (Color online) 
(a) $K=10$: $L$ dependence of the variance; (b) $K=0$: $L$-dependence of the variance; (c) $K=0$: $L$-dependence of the $Q$-variance.  The $Q$-variance for continuous distributions is defined as $\langle y^2 \rangle_Q \equiv \frac{\int_{-\infty}^\infty dy\,y^2[p_Q(y;\beta_Q)]^Q}{\int_{-\infty}^\infty dy\,[p_Q(y;\beta_Q)]^Q}$  (see details in \cite{TsallisMendesPlastino1998,FerriMartinezPlastino2005,Tsallisbook}). For $Q$-Gaussians, it is strightforwardly obtained $\beta_Q \langle y^2 \rangle_Q=\frac{1}{3-Q}$ (see, for instance, \cite{ThistletonMarshNelsonTsallis2007}). The green dashed lines indicate the analytical results, namely $1/2$ for $Q=1$ ($K=10$), and $0.939$ for $Q=1.935$ ($K=0$).
}}
\end{figure}

The numerical procedure for the CLT results can be summarized as follows: we firstly need to generate a long time 
series for the random variable of the system under consideration. This random variable is given by 
Eq.~(\ref{eq:yN}) for the standard map and by Eq.~(\ref{deltas}) for all the other models. 
The size of the time series is $5\times 10^7$ for the standard map and even larger than that for the others. 
Finally, the summation in Eq.~(\ref{eq:yN}) must be performed with several different values for $N$. 
After having enough data, one can generate the histograms for each case (with different $N$ values) using an 
appropriate box size allowing also a fair sampling of the central part. Two typical results obtained by this 
procedure are shown in Fig.~\ref{fig:Fig1}a and Fig.~\ref{fig:Fig1}c for the standard map.  \\

At this point, we analyze the large deviation behavior of the system, i.e., the speed of 
convergence of $Y_N/N$ for $N\to \infty$, by numerically exhibiting 
\begin{equation}
  P_N(Y_N/N > z) \approx  C(z) \,e^{-r_1(z) N} 
\end{equation}
with the {\it rate function} $r_1(z)$ defined by
\begin{equation}
 \label{r} 
 \lim_{N\to\infty} \frac{1}{N} \ln [P_N(Y_N/N > z) / C(z)] = -r_1(z) \;\;\;\;(r_1(0)=0)\, ,
\end{equation}
where $C(z)$ is a factor which, by definition, yields $C(z=0)=1/2$. 
Eq.~(\ref{r}) can be checked for $K=10$: see Fig.~\ref{fig:Fig1}c. 
The exponential decay is evident in the figure, from where the rate function is calculated. 
\noindent
Let us focus on the $K=0$ case. For this case, the random variables that we have defined previously 
using Eq.~(\ref{eq:yN}) are not anymore close to i.i.d., but strongly correlated instead. 
Therefore, the usual CLT cannot apply here and this constitutes a typical example to test the $Q$-CLT. 
If this kind of $Q$-generalised CLT holds, the sum of the correlated random variables is expected to converge to 
a $Q$-Gaussian, which is defined as:
\begin{equation}
 \label{q-Gauss} 
p_Q(y;\beta_Q) = A_Q\sqrt{\beta_Q} \, 
\exp_Q{\left[-\beta_Q \,y^2\right]} \;\;\;(\beta_Q>0) \, ,
\label{QGaussian}
\end{equation}
since this distribution optimises, under simple constraints, the continuous entropic form 
$S_q=k\frac{1-\int dx\,[p(x)]^q }{q-1}$ with $S_1=S_{BG}\equiv-k\int dx\,p(x) \ln p(x)$; 
the $Q$-Gaussian distribution is normalized for $Q<3$, and its second moment is finite for $Q<5/3$. 

Here, $1/\sqrt{\beta_Q}$ characterises the distribution width of $p_Q(y;B_Q)$, $A_Q$ being the normalisation factor: 
\begin{equation}
 \label{Aq}
 A_Q = \left\{\begin{array}{lc}\displaystyle\frac{\Gamma
 \left[\frac{5-3Q}{2(1-Q)}\right]}
 {\Gamma\left[\frac{2-Q}{1-Q}\right]}\sqrt{\frac{1-Q}{\pi}}  &if\;\; Q<1 \,,\\
  \displaystyle\frac{1}{\sqrt{ \pi }} &if \;\; Q=1 \,,\\
\displaystyle\frac{\Gamma \left[\frac{1}{Q-1}\right]}
{\Gamma\left[\frac{3-Q}{2(Q-1)}\right]}\sqrt{\frac{Q-1}{\pi}} \;\;\;\;\;  &if\;\; 1<Q<3 \,.
\end{array} \right.
\end{equation}
Note that, as $Q\to 1$, we recover the Gaussian distribution with the density given in Eq.~(\ref{density}). 
(For completeness, let us mention that it is claimed in \cite{Abe2010,Abe2010b} that the rather 
standard mathematical connection between the discrete and the continuous forms of $S_q$ carries some odd peculiarities. 
These claims have, however, been severely counter argued in \cite{Andresen2010,PlastinoRocca2017}).
Recent works \cite{ugur16} provide strong numerical evidence that, in this case, such a generalised 
central limit theorem appears to hold where the sequence $N^{\gamma}(Y_N/N)$ converges to a 
variable with a $Q$-Gaussian distribution for $N\to\infty$: 
\begin{equation}
\lim_{N\rightarrow\infty} P_N\left(N^{\gamma}Y_N/N > z\right) 
=\int_z^{\infty} p_Q(y;\beta_Q) dy \, .
\end{equation} 
For this to hold, we have to choose $Q \simeq1.935$ \cite{ugur16} and $\gamma \simeq0.65$ 
(adjusted from the best fit of the data); the exponent $\gamma$ plays, for $K=0$, the role of $1/2$ for $K=10$. 
A sufficient, but not necessary, condition for such a generalised central limit theorem is given by 
$Q$-independence \cite{UmarovTsallisSteinberg2008}. 
Note that, within the context of our example for $K=0$, $Q$-independence is not satisfied. 
Indeed, our $\gamma$ differs from $1/(4-2Q)$, the exponent which corresponds to the case of 
$Q$-independence. However, the probability density function is very well approximated by a 
$Q$-Gaussian with $Q \simeq 1.935$ as can be seen in Fig.~\ref{fig:Fig1}b. 
Let us emphasize at this point that $Q \simeq 1.935$ is the present numerical approximation 
for the recently established analytical results $Q=2$ \cite{BountisVeermanVivaldi2020}. 
For $K=0$, the problem of slow convergence of numerics to the analytical $Q=2$ result is due to strong 
correlations among the various initial conditions and one would need extremely large values of $N$ as well as 
of the number of initial conditions to tackle this. Obviously, this is not so for the uncorrelated case 
($K=10$) where numerical results approach the analytical ones rapidly. 

Now we are in the position of studying the speed of convergence of $Y_N/N$ for $N\to \infty$, 
as we have already done for $K=10$ case, thereby providing a large deviation analysis consistent 
with nonextensive statistical mechanics. Similarly, in this case, we provide numerical evidence for
a power-law emerging asymptotically for large values of $N$. Moreover, given the numerous works mentioned 
above that provide support to the field of $q$-statistics, and very especially in the LDT case 
in \cite{RuizTsallis2013}, we believe that the following form is particularly distinguished:   
\begin{equation}
  P_N(Y_N/N > z) \approx  C_q(z) \,e_q^{-r_q(z) N} \,,
\end{equation}
in the sense that an unique value $q$ might exist such that 
\begin{equation}
 \lim_{N\to\infty} \frac{1}{N} \ln_q [P_N(Y_N/N > z) / C_q(z)] = -r_q(z) 
\end{equation}
where the {\it $q$-rate function} $r_q(z) \ge 0$, the equality holding for $z=0$. 
We naturaly expect $C_q(z=0)=1/2$. 
In the $q\rightarrow 1$ limit, we recover $r_1(z)$ for $K=10$, as expected. 

Typical results are given in Fig.~\ref{fig:Fig1}d. The behavior is analogous to that of the usual case, 
the exponential being replaced by a $q$-exponential with $q=2.0$: compare Fig.~\ref{fig:Fig1}b of the 
usual LDT with Fig.~\ref{fig:Fig1}d for $q$-LDT. For $r_q(z)$, a numerical error up to 
$(q-1)|\ln_q(2P_{max}(z=0))|$ occurs, which vanishes for $q=1$,  $\forall P_{max}(z=0)$, 
and for $2P_{max}(z=0)=1$,  $\forall q$. In practice, for the present data it does not appear to overcome $1\%$. 
The $N$-dependence of the data-fit ratios of $P_N$ are depicted 
in Figs. \ref{fig:Fig1}e and \ref{fig:Fig1}f.  
The heights of the transients are of the same order (roughly 1.6 for typical values of $z$) for $K=10$ and $K=0$, 
but their durations are sensibly different. Indeed, we verify that the exponential behavior is attained, 
below a $5\%$ discrepancy, beyond $N \simeq 3 \times 10^6$ for $K=10$, whereas the possible $q$-exponential 
behavior is attained beyond $N \simeq 2 \times 10^5$ for $K=0$, i.e., 15 times earlier.

As a consistency check we have also calculated the variances corresponding to the $K=10$ and $K=0$ cases. 
For $K=10$ we calculated, from the computational discrete data, the second moment $\langle y^2 \rangle$ for 
increasing values of the cutoff $L \in [0,1/2]$ such that the data that are taken into account are those corresponding 
to $y/N \le L$, and compared it with the analytical expression obtained from the continuous Gaussian: 
see Fig.~\ref{fig:Figplus} (a). For $K=0$, we calculated, from the computational discrete data, the second moment 
$\langle y^2 \rangle$  as well as the $Q$-second moment $\langle y^2 \rangle_Q$ with $Q=1.935$, and compared the 
latter with the analytical expression obtained from the continuous $Q$-Gaussian in Eq.~(\ref{QGaussian}): 
see Figs. \ref{fig:Figplus} (b) and (c). Naturally, for $K=0$ the $L$-dependences of  
$\langle y^2 \rangle$ and $\langle y^2 \rangle_Q$ differ dramatically. Indeed, since $Q=1.935 >5/3$, 
the standard variance is mathematically ill-defined (see, for instance, \cite{ThistletonMarshNelsonTsallis2007}).

\begin{figure}
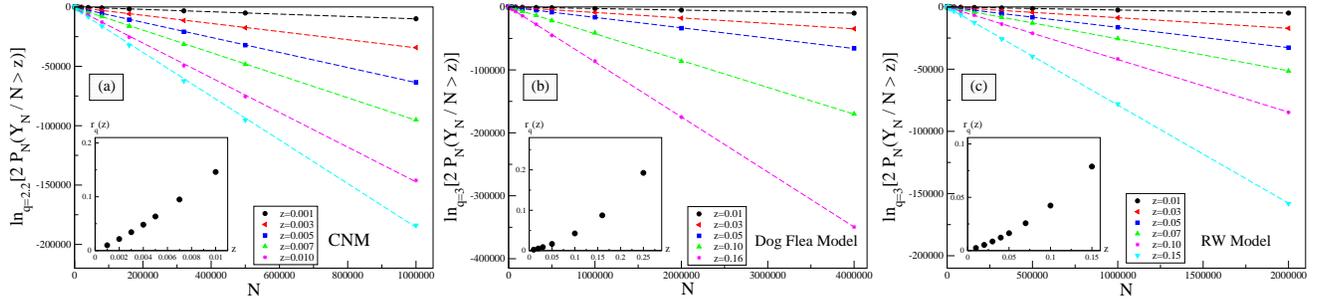
 
  \centering
  \includegraphics[scale=0.23,angle=0]{fig3a.eps}
   \includegraphics[scale=0.23,angle=0]{fig3b.eps}
    \includegraphics[scale=0.23,angle=0]{fig3c.eps}
  \caption{\label{fig:Fig3} {\small (Color online) 
  Large deviation probability $P_N(Y_N/N > z)$ as a function of $N$ ($q$-log - linear representation) 
  for (a)~the CNM with $\sigma=0.05$ ($Q\simeq 2.1$ and $q\simeq 2.2$), 
  (b)~the Ehrenfest dog flea model ($Q\simeq 2.32$ and $q\simeq 3$), and 
  (c)~the random walk avalanche model ($Q\simeq 2.32$ and $q\simeq 3$). 
  The behavior of $r_q(z)$ can be seen in the insets of each figure. 
  We remind that straight lines in $q$-log - linear representation strictly guarantee a 
  $q$-exponential behavior only if verified at {\it all} scales. Otherwise, they only guarantee an asymptotic 
  power-law behavior for large values of $N$. A detailed illustration of this point is presented 
  in Fig.~\ref{fig:Fig1}.
  }}
  \end{figure}

We address next the return distributions of some paradigmatic models, namely, 
the {\it Coherent Noise Model} (CNM), the {\it Ehrenfest Dog Flea Model} (EDFM), and the 
{\it Random Walk Avalanches Model} (RWAM). The CNM has been introduced firstly for analyzing 
biological extinctions \cite{CNM}, but then it has been also used as a simple mean-field model 
for earthquakes \cite{CNM2}. 
The system consists of $N$ agents, each of which having a threshold $x_i$ against an external 
stress $\eta$. In the model, these parameter values are chosen randomly from probability 
distributions $p_{thresh}(x)$ and $p_{stress}(\eta)$, respectively. 
Generically, exponential distribution $p_{stress}(\eta)=(1/\sigma)\exp(-\eta/\sigma)$ is used for the 
external stress, whereas, for  $p_{thresh}(x)$, the uniform distribution ($0\leq x\leq 1$) is chosen. 
The dynamics of the model can be given in three steps: 
(i)~a random stress $\eta$ is generated from $p_{stress}(\eta)$ and all agents with $x_i\leq\eta$ 
are replaced by new agents with new threshold drawn from $p_{thresh}(x)$, 
(ii)~finally, a small fraction $f$ of $N$ agents is chosen and new thresholds drawn from 
$p_{thresh}(x)$ are assigned, then 
(iii)~these steps are used for the next time step. 
The number of agents replaced in the first step determines the event size $s$ \cite{CNM,CNM2}.
The return distribution is noted $p(\Delta s)$, where 
\begin{equation}
\Delta s = s(t+1) - s(t)
\label{deltas}
\end{equation} 
is the difference between two consecutive event sizes. 
This quantity here plays the role of the random variable $Y_N$ that we have used for the 
standard map. 
The corresponding distributions were studied 
in \cite{CelikogluTirnakliQueiros2010,ChristopoulosSarlis2014,ChristopoulosSarlis2017}. 
The return distributions appear to follow the $Q$-Gaussian form. 
For example, for  $\sigma=0.05$  it is $Q \simeq 2.10$ \cite{CelikogluTirnakliQueiros2010}. 
We verify here that the large deviation probability decays as a power-law which is consistent with a
$q$-exponential with $q \simeq 2.20$ (see Fig.~\ref{fig:Fig3}a).  

The Ehrenfest Dog Flea Model (EDFM) has been introduced in 1907 by Ehrenfest and Ehrenfest \cite{Ehrenfest}. 
It is a simple and paradigmatic model of generation-recombination Markov chain describing the 
process of approaching an equilibrium state in a large set of uncoupled two state systems 
together with fluctuations avalanches around this state \cite{DF}.
The simple dynamics of the model is the following: It has $N$ dynamical sites represented by 
the total number of fleas shared by two dogs, namely dog $A$ and dog $B$. Suppose that there are 
$N_{A}$ fleas on $A$ and $N_{B}$ fleas on $B$ which leads to the population $N=N_{A}+N_{B}$. 
At every time step, a randomly chosen flea jumps from one dog to the other which results in 
$N_{A}\rightarrow N_{A}\pm1$ and $N_{B}\rightarrow N_{B}\mp1$. This procedure is repeated for 
an arbitrary number of times. In the long run, the mean number of fleas on both $A$ and $B$ 
converges to the equilibrium value, $\langle N_{A}\rangle =\langle N_{B}\rangle = N/2$ with some 
fluctuations. A single fluctuation is described as a process that starts once the number of 
fleas on one of the dogs becomes larger (or smaller) than the equilibrium value $N/2$ and stops 
when it gets back to it for the first time. Therefore, termination of a fluctuation specifies 
the start of the other one. The length ($\lambda$) of a fluctuation is determined by the number 
of time steps elapsed until the fluctuation ends.    
The return distribution is defined as in the CNM through Eq.~(\ref{deltas}). 
It was analysed in \cite{BakarTirnakli1} and it obeys a $Q$-Gaussian form with $Q\simeq 2.32$. 
We found that the large deviation probability decays 
as a power-law which is consistent with a $q$-exponential with $q=3$, 
as seen in Fig.~\ref{fig:Fig3}b.  


It was shown in the RWAM \cite{Kim} that the avalanche size of a one-dimensional directed 
sandpile model can be mapped to the area under a Brownian curve with an absorbing boundary at the origin. 
This is equivalent to a random walker on $[0,\infty)$ with an absorbing boundary at the origin. 
If we denote the trajectory of the random walker by $x(i)$ with $i=0,1,...,N$, the avalanche 
size can be described as $s = \sum_{i=1}^N x(i)$ ($x(0)=1$).
The return distribution of this model is defined as in the CNM, i.e., using Eq.~(\ref{deltas}), 
and it was numerically found a $Q$-Gaussian with $Q \simeq 2.32$ \cite{Tirnakliunpublished}. 
We have determined here that the large deviation probability decays as a power-law consistent with a
$q$-exponential with $q \simeq 3.0$, as seen in Fig.~\ref{fig:Fig3}c.


Finally let us summarize the numerical procedure for the CLT and LDT results. We firstly need to generate a 
long time series for the random variable of the system under consideration. This random variable is given by 
Eq.~(\ref{eq:yN}) for the standard map and by Eq.~(\ref{deltas}) for all the other models. 
The size of the time series is $5\times 10^7$ for the standard map and even larger than that for the others. 
Finally, the summation in Eq.~(\ref{eq:yN}) must be performed with several different values for $N$. 
After having enough data, one can generate the histograms for each case (with different $N$ values) using an 
appropriate box size allowing also a fair sampling of the central part. Two typical results obtained by this 
procedure are shown in Fig.~\ref{fig:Fig1}a and Fig.~\ref{fig:Fig1}c for the standard map. 
To achieve the large deviation analysis, firstly one needs to localize an $x$-axis value in the histogram, 
denoted by $z$, as seen in Figs.~\ref{fig:Fig1}a and \ref{fig:Fig1}b. Then, we calculate the total probability 
value larger than this $z$. This allows us to construct a plot that represents this value with respect to various 
$N$ for any particular $z$ value chosen. This can be seen in Figs.~\ref{fig:Fig1}b and \ref{fig:Fig1}d 
for the standard map and in Fig.~\ref{fig:Fig3} for the other models.

\section{Discussion}
Let us conclude by reminding that our aim is to approach the fingerprints of 
Boltzmann-Gibbs statistical mechanics, namely the Maxwellian distribution of velocities and 
the BG exponential weight for the energies, within a more general context. Indeed, in the 
realm of nonextensive statistical mechanics based on nonadditive entropies, a $Q$-Gaussian 
distribution emerges for the velocities and a $q$-exponential weight emerges for the energies, 
with $(Q,q) \ne (1,1)$, the equality $Q=q=1$
holding precisely for the BG theory. 
These generalised results appear to respectively follow along the lines of 
the Central Limit Theorem and the Large Deviation Theory. This scenario has already been 
successfully verified for a purely probabilistic model (see \cite{RuizTsallis2013} and references therein). 
We numerically checked here this conjectural path with iterative dynamical models, namely 
the paradigmatic two-dimensional conservative standard map,
the coherent noise model, and the random walk avalanches, as well as a genuine $N$-body system, the Ehrenfest 
dog-flea model.
In all cases the conjecture appears to lurk.
Notice however that, for the standard map, we define sums 
of $N$ terms precisely as in the Central Limit Theorem. For the other three illustrations we simply use 
the return distributions, which in all cases are of the $Q$-Gaussian form. In the present four examples, 
the Large Deviation probability asymptotically decays with $N$ as a power-law which appears to correspond to a 
$q$-exponential form with the argument $r_qN$ being proportional to $N$, which would in turn be consistent with 
the Legendre structure of thermodynamics. Indeed, the intensive quantity $r_q$ possibly is related to a nonadditive 
relative entropy {\it per particle}. 
Along the lines of the present promising results, analytical approaches would naturally be very welcome, 
either for the specific models studied here, or in the ambitious form of a $q$-generalised theorem for 
large deviations based on a $Q$-generalised central limit theorem for strongly correlated random variables.
To be more precise, we generically expect that many complex systems would exhibit an asymptotic 
power-law at their large deviation behaviour. On the other hand, whenever the central limit behavior is concerned, 
many systems present $Q$-Gaussians as attractors and we then consistently expect $q$-exponentials for the 
large deviation probabilities with a value for $q$ which univocally depends on $Q$, and which satisfies $q=Q=1$. 
It would certainly be 
interesting that future analytical and/or numerical efforts would focus along these lines for typical 
complex natural, artificial and social systems. It is important to have in mind that, whereas the 
asymptotic power-law behavior appears naturally for large values of $N$ (as shown here), the establishment of the 
distinctive $q$-exponential form demands the exploration along {\it all} scales for $N$, particularly those 
involving relatively small values of $N$.  
Let us nevertheless emphasise that the ubiquity of these power-laws (here and in Ref. \cite{RuizTsallis2013}) 
calls for a sensible generalisation of the standard LDT exponential behavior. An admissible such generalisation 
should in principle satisfy two conditions: (i) to asymptotically exhibit, of course, the power-law, and (ii) 
its argument should be extensive, i.e., proportional to $N$. The $q$-exponential possibility with $q>1$ 
satisfies both.

\section{Acknowledgements}
We are thankful to two anonymous Referees for useful remarks. Also, we acknowledge partial financial support 
by the Santa Fe Institute, New Mexico, and the Max Planck Institute for Mathematics in the Sciences, Leipzig, 
Germany, for support of the SFI Micro Working Group “Large Deviations in Complex Systems” meeting. 
CT acknowledges as well CNPq and Faperj (Brazilian agencies). 
U.T. is a member of the Science Academy, Bilim Akademisi, Turkey and acknowledges partial support from 
TUBITAK (Turkish Agency) under the Research Project number 121F269.

%



\end{document}